\documentstyle[prb,preprint,epsfig,aps]{revtex}
\tightenlines
\begin{document}
\draft

\title{Optical conductivity of the non-superconducting cuprate 
La$_{8-x}$Sr$_x$Cu$_8$O$_{20}$}
\author{A. Lucarelli, S. Lupi, P. Calvani, and P. Maselli} 
\address{Istituto Nazionale di Fisica della Materia - Dipartimento 
di Fisica, Universit\`a di Roma La Sapienza, Piazzale Aldo Moro 2, 
I-00185 Roma, Italy}
\author{G. De Marzi and P. Roy}
\address{Laboratoire pour l'Utilisation du Rayonnement
Electromagn\'etique, Universit\'e Paris-Sud, 91405 Orsay, France} 
\author{N. L. Saini and A. Bianconi}
\address{Istituto Nazionale di Fisica della Materia - Dipartimento 
di Fisica, Universit\`a di Roma La Sapienza, Piazzale Aldo Moro 2, 
I-00185 Roma, Italy}
\author{T. Ito and K. Oka}
\address{National Institute of Advanced Industrial Science and 
Technology, 1-1-1 Umezono, Tsukuba 305-8568, Japan}

\date{\today}
\maketitle

\begin{abstract}
La$_{8-x}$Sr$_x$Cu$_8$O$_{20}$ is a non-superconducting cuprate 
which exhibits a doubling of the elementary cell along the $c$ 
axis. Its optical conductivity has been first 
measured here, down 
to 20 K, in two single crystals with $x$ = 1.56 and $x$ =  2.24. 
Along $c$, bands are observed in both samples which
correspond to strongly bound charges, and confirm that the 
cell doubling is due to charge ordering. In the $ab$ plane, in 
addition to the Drude term one observes an infrared peak at $\sim$ 
0.1 eV and a midinfrared band 
at 0.7 eV. The 0.1 eV peak is found at higher frequencies below 200 
K, in correspondence of an anomalous increase in the dc 
resistivity and consistently with its assignment to localized 
charges. These results point out similarities and differences with 
respect to the optical properties of superconducting cuprates.
\end{abstract} 
\narrowtext

\section{Introduction}

In recent years, several studies have been devoted to the problem 
of localization and ordering of the doped charges in the cuprates. 
Theories of their metallic phase in terms of fluctuating charged 
stripes\cite{Emery,Castellani,Salkola,Grilli} have been proposed. 
Models of high-$T_c$ 
superconductivity\cite{Ranninger,Bianconi,Quemerais} have also 
been proposed, which assume a coexistence of free and bound 
charges. Experimentally, charged superlattices have been clearly 
detected by diffraction techniques in compounds\cite{Chen}  like 
La$_{2-x}$Sr$_x$NiO$_4$ or La$_{2-x}$Sr$_x$MnO$_4$, that are 
isostructural to La$_{2-x}$Sr$_x$CuO$_4$ (2-1-4). In this latter 
however, ordered arrays of spin and charge
have been observed only upon partial replacement of La by 
Nd.\cite{Tranquada} On the other hand, charged stripe fluctuations 
have been detected in YBaCu$_3$O$_{7-x}$ by use of neutron 
scattering\cite{Mook} and ion channeling\cite{Sharma} as well as, 
in 2-1-4, by X-ray absorption\cite{Bianconi96} and angle-resolved 
photoemission.\cite{Ino} 

As far as the optical spectra are concerned, it is well known that 
the complex optical conductivity $\tilde \sigma (\omega)$ of a 
metallic cuprate, measured in the $ab$-plane, does not follow 
a normal Drude behavior. In the 
literature, this effect has been  taken into account by using two 
different approaches.
According to the so-called "one component", or "anomalous-Drude" 
model,\cite{Schlesinger,Timusk} one assumes that
both the carrier relaxation time $\tau$ and its effective mass
$m^*$ are functions of the photon frequency $\omega$: 

\begin{equation}
\tilde\sigma(\omega) = {ne^2\tau(\omega) \over m^*(\omega)[1-i\omega
\tau(\omega)]}
\label{anomalous}
\end{equation}

\noindent
This approach, that is most suitable to fit the smooth spectra of 
the HCTS close to optimum doping, has also been employed 
to extract from $\tilde \sigma (\omega)$ optical 
pseudogaps in a variety of cuprates.\cite{Timusk}. 

According to an alternative, multi-component model, the real part of the infrared conductivity 

\begin{equation}
\sigma (\omega) = {\omega \over {4 \pi}} Im [{\tilde \epsilon}] 
\, ,
\label{sigma-ep}
\end{equation}

\noindent
can be fitted by a Drude-Lorentz dielectric function

\begin{equation}
\tilde \epsilon = \epsilon_{\infty} - {\omega_D^2 \over {\omega^2 
- i \omega \Gamma_D}} + \sum_j {S_j^2 \over {(\omega^2 - \omega_j^2) 
- i\omega \Gamma_j}}\,  .
\label{multicomp}
\end{equation}

\noindent
where, in the sum, one oscillator accounts for the so-called $d$ 
band, another one for the midinfrared (MIR) band. Two further 
oscillators are often needed to 
reproduce the Cu-O charge-transfer band which appears in the near-
infrared and the visible.\cite{prb92} The analysis of $\sigma 
(\omega)$ in terms of Eq.\ \ref{multicomp} is supported by the fact 
that the $d$ and MIR contributions, which are needed to fit the 
spectra at optimum doping, have been directly resolved in a number 
of cuprates, both insulating and metallic. 

The $d$-band is observed at $\omega_d \sim$ 0.1 eV in both lightly 
doped La$_2$CuO$_{4+y}$,\cite{Falck} and Nd$_2$CuO$_{4-
y}$,\cite{prb96,Thomas} and is extremely sensitive to both 
doping and temperature. In Nd$_{2-x}$Ce$_x$CuO$_{4-
y}$,\cite{prl99} it increases in intensity and displaces towards 
lower energies for both increasing doping and lowering 
temperature.\cite{prl99} The 
softening of an infrared band for $T \to 0$ is quite unusual as it 
will be shown here also, and allows one to make interesting 
comparisons with the non-superconducting oxides. The $d$ contribution 
is still needed to fit $\sigma (\omega)$ at $x \simeq 0.15$, 
where Nd$_{2-x}$Ce$_x$CuO$_{4-y}$ is superconducting at optimum doping. 
Therein, it is found at frequencies even lower than 
the softest transverse phonon mode.\cite{prl99,Basov01} Finally, 
it disappears in the normal metallic phase at high doping ($x > 
0.18$). Its presence in superconducting 
cuprates, at least those with low $T_c$, is confirmed by recent 
observations on superconducting Bi$_2$Sr$_2$CuO$_6$ close to 
optimum doping ($T_c$ = 20 K).\cite{prb00} 

The above observations are explained by assuming that the $d$ band 
is due to polaronic charges, that increasingly selftrap at low $T$ 
due to the competition between thermal excitations and charge-
lattice interaction.\cite{prb96,Falck,prl99,Salje,Lobo}. The 
softening of the charge binding energy for increasing polaron 
density (i.e. for increasing doping and/or decreasing 
temperature) is explained in terms of polaron-polaron 
interactions.\cite{Fratini,Cataudella,Tempere1,Tempere2} 
The strength of the $d$ band seems also to increase as the 
dimensionality of the environment  decreases. In 
YBa$_2$Cu$_4$O$_8$, an untwinned cuprate of the YBCO family that 
has both conducting Cu-O planes and Cu-O chains, the optical 
conductivity shows a Drude contribution well distinguished from a 
huge polaronic peak at 0.1 eV, when the radiation field is 
directed along the chains.\cite{Bucher}
In turn, the nearly $T$-independent midinfrared band 
(MIR) has been observed both in layered and cubic perovskites, 
upon doping, at $\approx$ 0.5 eV.\cite{prb92,Uchida,Blanton} This 
band, which close to the MIT transition helps to build up the 
Drude term with part of its spectral 
weight,\cite{prb92,Uchida,Blanton} is usually attributed to 
electronic states created by doping in the Cu-O charge-transfer 
gap. 

In the present paper the optical properties of La$_{8-
x}$Sr$_x$Cu$_8$O$_{20}$ (8-8-20) will be studied, and analyzed  by 
using the model of Eq.\ \ref{multicomp}. La$_{8-
x}$Sr$_x$Cu$_8$O$_{20}$ contains the same 
chemical species as the 40 K superconductor 2-1-4. However, i) it 
is not superconducting for any $x$ ; ii) it exhibits in the 
electron diffraction spectra well defined superlattice spots for 
$x \sim 1.6$, indicating unit-cell doubling along the $c$ axis,
diffused spots for $ \sim 2.2$.\cite{Yamaguchi} Therefore, an 
infrared study of 8-8-20 can add information on the ordering 
process that takes place in this cuprate and provide interesting 
comparisons with the optical behavior of the superconducting 
cuprates.

\section{Sample description and experimental procedure}
 
The crystal strucuture of La$_{8-x}$Sr$_x$Cu$_8$O$_{20}$ is 
tetragonal,\cite{Rakho} with lattice constants $a_0$ = $b_0$ = 
1.084 nm, $c_0$ = 0.3861 nm. It can be derived from that of 
La$_{2-x}$Sr$_x$CuO$_4$ (2-1-4) by eliminating oxygen ions in a 
regular way. As a result, one is left with corner-sharing Cu-O$_6$ 
octahedra, Cu-O$_5$ pyramids and Cu-O$_4$ squares. 
The latter ones form one-dimensional (1D) chains along the $c$ axis, 
while the corner-sharing Cu-O$_6$ octahedra and the Cu-O$_5$ 
pyramids form a three-dimensional (3D) network of essentially 1D 
paths. The charges travel along this network and there are no 
Cu-O conducting sheets. 
The transport properties of La$_{8-x}$Sr$_x$Cu$_8$O$_{20}$ have 
been investigated on single crystals with $1.56 < x <2.24$. In 
this range the nominal charge varies from 0.2 to 0.3 holes per Cu 
atom, compared with 0.06 to 0.3 holes per Cu atom in the metallic 
phase of La$_{2-x}$Sr$_x$CuO$_4$. Resistivity measurements on 
single crystals showed that 8-8-20, similarly to other cuprates, 
has an "anomalous" metallic region for $1.5 \alt x \alt 1.8$, 
followed by a normal metallic phase for $x \agt 2$; however, for 
any $x$, it is 
not superconducting down to 1.3 K.\cite{Ito99} At room temperature 
the anisotropy in the resistivity is $\rho_c/\rho_{ab} \sim$ 10 
at any $x$. At constant temperature, both $\rho_{ab}$ and 
$\rho_{c}$ decrease for increasing $x$, as usually observed in 
doped Mott insulators. As a function of $T$, both 
$\rho_{ab}$ and $\rho_{c}$ are metallic-like in the whole doping 
range, but exhibit an anomalous broad maximum\cite{Ito97} 
between two temperatures 
$T_{c1}$ and $T_{c2}$  (with $T_{c1} > T_{c2}$) which change with 
$x$. $T_{c1}$  has been associated with a reduction in the 
scattering rate of the itinerant holes, related to a weak 
ferromagnetic ordering in the 3D network. Below $T_{c2}$  an 
antiferromagnetic (AFM) order is observed, and attributed to the 
chains of Cu-O$_4$ squares aligned along the $c$ axis.\cite{Ito99} 
The AFM transition is associated with strong, opposite variations 
of the Hall coefficients $R_H$ in the $ab$ plane and along the $c$
axis, and also with a sudden change in $\rho_{ab}$. These 
effects have been explained with the formation of a gap at the 
Fermi surface along certain directions, due to the formation of 
spin density waves at $T_{c2}$.\cite{Ito97} Both $T_{c1}$ and 
$T_{c2}$ decrease by increasing $x$, until a conventional metallic 
behavior is established in the sample with x=2.24.\cite{Ito97} As 
already mentioned, electron diffraction studies\cite{Ito99} have 
shown that an ordering process causes a doubling of the elementary 
cell along the $c$ direction. According to the authors, the 
ordering involves the charges introduced by doping, most probably 
in the CuO$_4$ squares. Indeed, one may remark that the anomalies 
in the resistivity of 8-8-20 are quite similar to those detected 
in compounds like NbSe$_3$, where one-dimensional charge density 
waves form below a critical temperature.\cite{Chaussy} 

The two single crystals selected for the present optical study of 
La$_{8-x}$Sr$_x$Cu$_8$O$_{20}$ have $x$ = 2.24 and 
$x$ = 1.56, the highest and nearly the lowest 
doping level, respectively, that have been studied in the 
literature. Basing on the transport 
properties\cite{Ito99} of crystals obtained from the same batch, 
the former should provide a good metallic spectrum for reference 
and the latter one, which has $T_{c1}$ = 145 K and $T_{c2}$ = 85 K, 
is 
expected to exhibit 'anomalous' spectral features in the clearest
way. Both crystals were grown by the travelling-
solvent floating zone method.\cite{Ito97} Their chemical 
composition was determined by the inductively coupled plasma 
atomic emission (ICP-AES). The transport and magnetic properties 
of the samples have been described in Ref. \onlinecite{Ito97}. 

The samples were mounted on the cold finger of a two stage closed-
cycle cryostat, whose temperature was kept constant within ±2 K 
and could be varied from 295 to 20 K. The reflectance $R(\omega)$ 
of the samples, with the radiation field in the $ab$ plane, was 
measured relative to gold- and aluminum-plated references. With 
the radiation field polarized along the $c$ axis, the reference 
was obtained by evaporating directly gold on the 
sample.\cite{Homes}   Thus we obtained reliable absolute values of 
the reflectivity in spite of the small transverse 
dimension of the sample. The spectra were collected 
by a rapid scanning interferometer, typically from 80 cm$^{-1}$ to 
20000 cm$^{-1}$. A 
Drude-Lorentz fit was used to extrapolate the reflectivity to 
$\omega = 0$. On the high-energy side, our data were extrapolated 
with the reflectivity reported in Ref. \onlinecite{Tajima} for
La$_{2-x}$Sr$_x$CuO$_4$, under the reasonable assumption that the 
ultraviolet bands of 8-8-20 are not too different from those of 
2-1-4. The real part of the optical conductivity $\sigma (\omega)$ 
was then extracted from $R(\omega)$ by usual Kramers-Kronig 
trasformations. 

\section{Results and discussion}

\subsection{Optical response of the $c$ axis}

\begin{figure}
{\hbox{\psfig{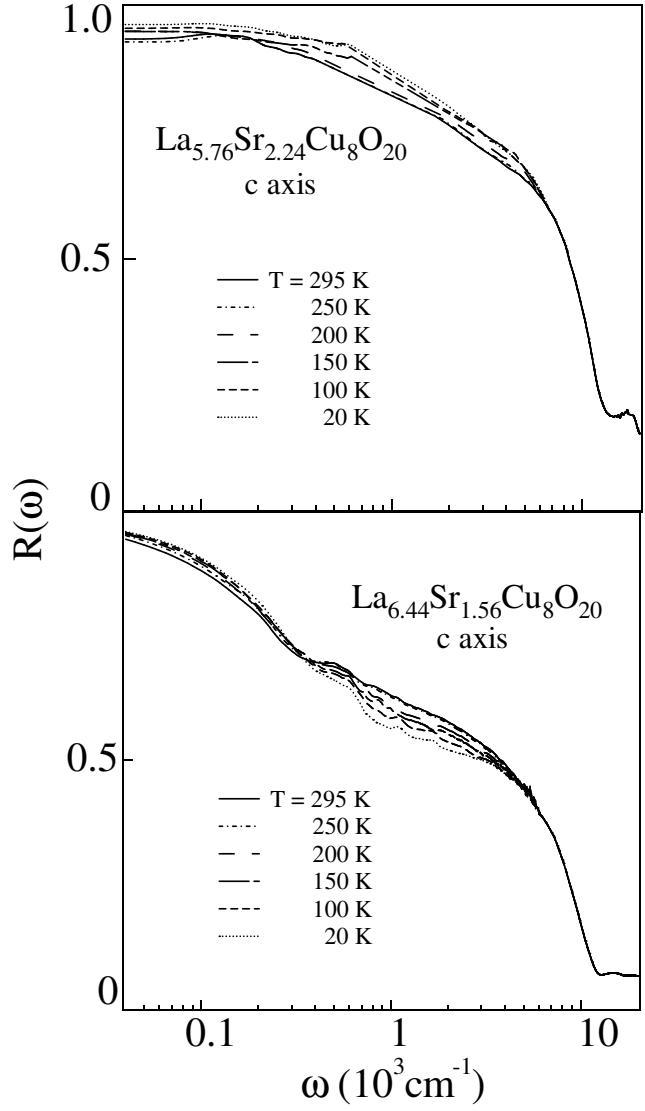}}}
\caption{Infrared reflectivity measured up to 
to 20000 cm$^{-1}$ at different temperatures, with the 
radiation field polarized along the $c$ axis for both $x$ = 2.24 
(top) and $x$ = 1.56 (bottom).}
\label{R-c}
\end{figure}

The reflectivity $R(\omega)$ measured along the $c$ axis of 
La$_{8-x}$Sr$_x$Cu$_8$O$_{20}$ is shown in Fig.\ \ref{R-c} for x = 
2.24 (top) and x = 1.56 (bottom). The spectra are reported in the 
range from 40 to 20000 cm$^{-1}$ at six different temperatures. 
$R(\omega)$ 
exhibits a metallic-like behavior in both samples with a well 
evident pseudo-plasma edge around 10000 cm$^{-1}$. The 
electronic band in the visible range is similar to features 
observed in most cuprates\cite{prb92,Uchida} and attributed to the
Cu-O charge-transfer transitions. For $x$ = 2.24, it can be 
reproduced by two Lorentzians placed at 17800 and 21200 cm$^{-1}$, 
with $\epsilon_{\infty}$ = 4.1. At  low frequency $R(\omega)$ 
differs in the two crystals. The 
sample with $x$ = 2.24 shows a standard metallic reflectivity, 
except for a smooth anomaly at $\sim$ 600 cm$^{-1}$ for $T \alt$ 
150 K. On the other hand, for $x$ = 1.56 there is a 
clear change of slope in $R(\omega)$  around 300 cm$^{-1}$, 
suggestive of two different contributions to $\sigma (\omega)$. 
The contribution at  high frequency exhibits a more pronounced 
temperature dependence than that at low frequency. 

\begin{figure}
{\hbox{\psfig{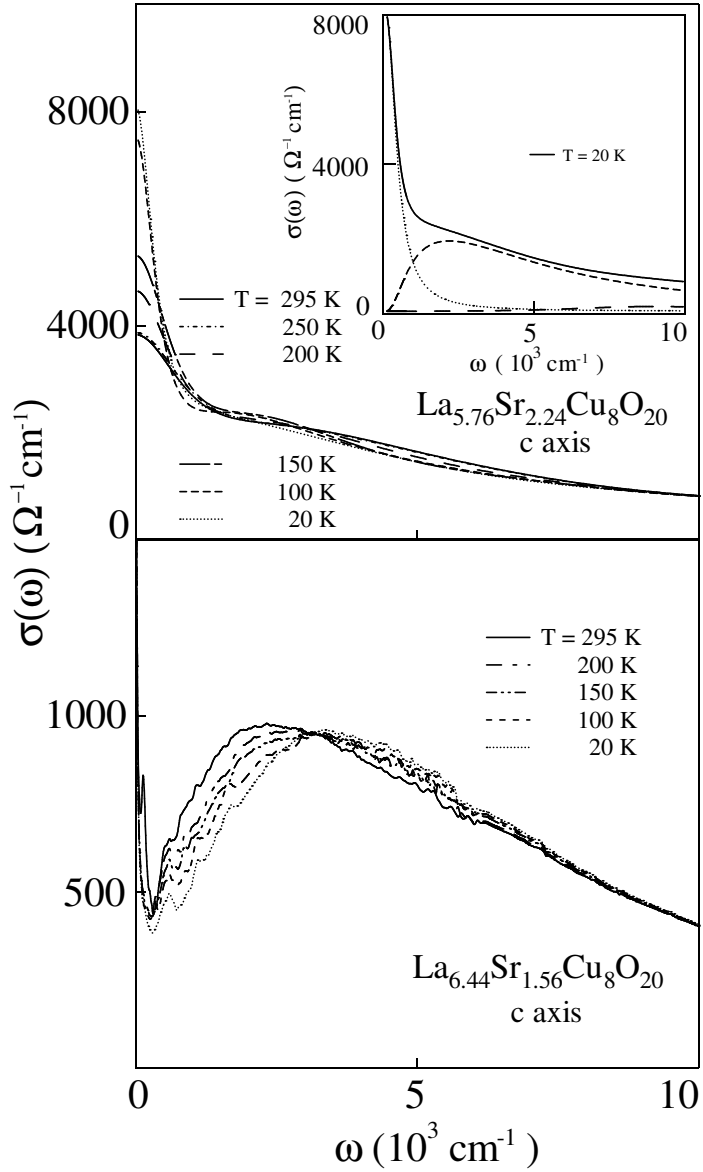}}}
\caption{Infrared optical conductivity of the $c$ axis at 
different temperatures, as extracted from the $R(\omega)$ of Fig. 
1, for $x$= 2.24 (top) and $x$ = 1.56 (bottom). The 
inset compares the experimental $\sigma(\omega)$ at 20 K (solid 
line) with a fit based on a conventional Drude term (dotted line) 
and two contributions at infrared frequencies (dashed lines).}
\label{sigma-c}
\end{figure}

The multi-component structure of the absorption is evident 
in the real part of the optical conductivity, reported for both 
crystals in Fig.\ \ref{sigma-c}. In the sample with $x$ = 2.24 
(top), $\sigma(\omega)$ exhibits two well-defined components in 
the infrared, with different temperature 
behaviors: a Drude term which is dominating for $\omega \alt$ 1500 
cm$^{-1}$ and narrows for $T \to 0$, and a broad band in the 
midinfrared. This latter is separated from the Drude contribution 
by a change of slope at $\sim$ 1500 cm$^{-1}$, more pronounced at 
low $T$. The inset compares the experimental $\sigma(\omega)$ at 
20 K with a fit to Eq.\  \ref{multicomp}. In the frequency range 
shown in the inset, this includes a Drude term with  $\omega_p 
\simeq $ 1500 cm$^{-1}$  at  all temperatures and a 
$\Gamma_D$ which decreases from 1200 cm$^{-1}$ at 295 K to 400 
cm$^{-1}$ at 20 K,  a  contribution in the midinfrared peaked at 
2300 cm$^{-1}$ at 20 K, and a weak background centered at about 
9000 cm$^{-1}$. Basing on these results, it seems quite reasonable 
to assign the midinfrared  band to those bound charges which 
produce a diffuse scattering in the electron diffraction spectra 
of the $x$ = 2.2.4 sample. In turn, the strong 
Drude term accounts for the good dc conductivity of this compound 
along the $c$ axis. Therefore the present data show a 
coexistence of free and bound charges in the cuprate even at $x$ = 
2.24, a doping value which provides at all temperatures the lowest 
dc resistivity reached by this compound.

A Drude-like absorption and a band in the midinfrared are found
also in the sample with $x$ = 1.56 (bottom panel of Fig.\ 
\ref{sigma-c}) where, however, they are directly resolved in the 
$\sigma(\omega)$. This is due to the fact that the Drude term is 
weaker than at $x$ = 2.24 by a factor of 10, the midinfrared band 
by a factor of 2. By recalling the results of ref. 
\onlinecite{Ito97}, one may 
assign the sharp, $T$-dependent midinfrared band here observed for 
$x$ = 1.56 to the photoexcitation of those bound charges which,  
for $x$ = 1.6, contribute sharp superlattice spots to the 
electron-diffraction spectra. The peak frequency increases 
monotonically from  
$\sim$ 2300 cm$^{-1}$ at 295 K to $\sim$ 3300 cm$^{-1}$ at 20 K. 
At the latter temperature, the corresponding band in the inset of 
Fig.\ \ref{sigma-c}) is peaked at $\sim$ 2000 cm$^{-1}$. As 
already mentioned, a softening of the bound-charge absorption for 
increasing doping has also been observed in superconducting 
families of cuprates.\cite{prl99} If one describes the bound 
charges in terms of small polarons, as previously done for other 
perovskites with charge ordering,\cite{Bi,rc96} the peak energy of 
the band is just twice the charge-lattice binding energy 
$E_p$.\cite{Emin} From Fig.\ \ref{sigma-c}, at $x$ =1.56  one thus 
finds $E_p \sim$ 1100 cm$^{-1}$ at 295 K, $E_p \sim$ 1600 
cm$^{-1}$ at 20 K.

The full opening of a charge-ordering gap in $\sigma(\omega)$, as 
observed  for 
instance\cite{rc96} in La$_{1.67}$Sr$_{0.33}$NiO$_4$, is prevented 
in La$_{6.44}$Sr$_{1.56}$Cu$_8$O$_{20}$ by the Drude-like 
component, which in Fig.\ \ref{sigma-c} is observed at any 
temperature. This shows that the bound charges are coexisting with 
free charges in the 1.56 crystal, consistently with its not 
negligible dc conductivity. Due to the peculiar structure of 
8-8-20 along the $c$ axis, the two types of charges are likely to be 
even spatially separated. The ordered charges can occupy the 
CuO$_4$ chains, as already proposed,\cite{Ito99} while the free 
carriers may travel along the zig-zag paths formed by the 
polyhedrical network, which are aligned in average along the $c$ 
direction. A similar situation is encountered for instance in 
NbSe$_3$, which remains 
metallic down to the lowest temperatures in spite of the formation 
of charge-density waves. These latters  appear in two out of the 
three infinite-length trigonal prisms that build up the structure 
of NbSe$_3$, while the third one remains metallic.\cite{Monceau}

\subsection{Optical response of the $ab$ plane}

\begin{figure}
{\hbox{\psfig{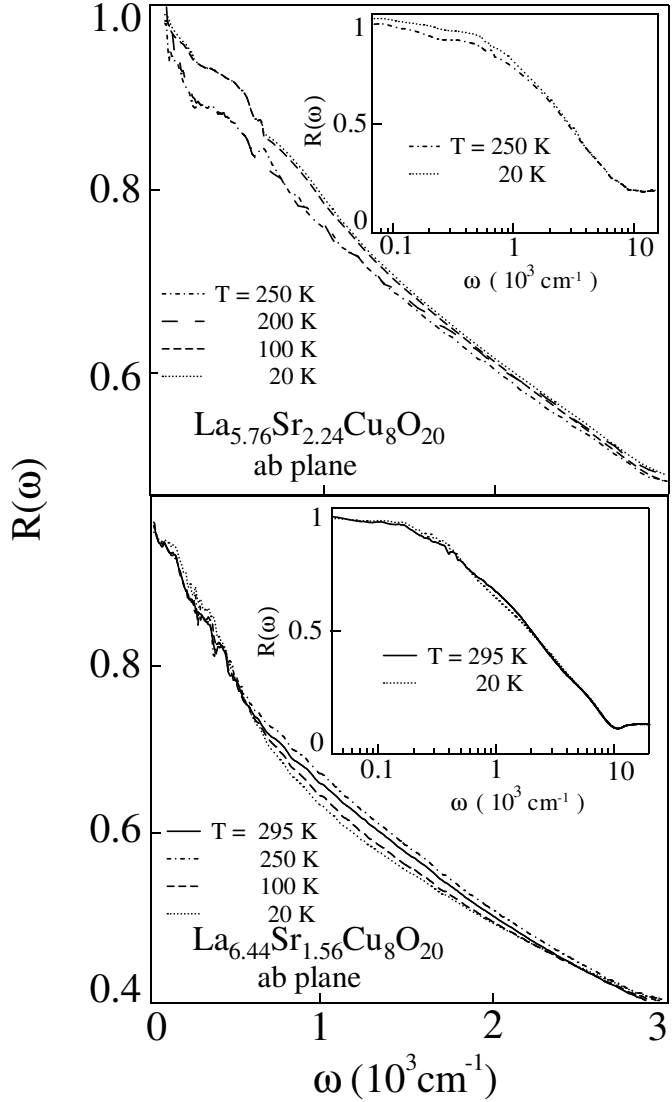}}}
\caption{Infrared reflectivity measured at different temperatures, 
with the radiation field polarized in the $ab$ plane for both $x$ 
= 2.24 (top), and $x$ = 1.56 (bottom). In the insets, $R(\omega)$ 
is shown at two temperatures up to 20000 cm$^{-1}$.}
\label{R-ab}
\end{figure}

The reflectivity $R(\omega)$ of the $ab$ plane of La$_{8-
x}$Sr$_x$Cu$_8$O$_{20}$ is shown in Fig.\ \ref{R-ab} for $x$ = 
2.24 (top) and $x$ = 1.56 (bottom). The spectra are reported in 
the range from 80 to 3000 cm$^{-1}$ at different temperatures. In 
the insets of both figures $R(\omega)$ is shown at two 
temperatures in the energy range from 80 to 20000 cm$^{-1}$. 
$R(\omega)$ presents a metallic-like behavior in both samples with 
a well evident pseudo-plasma edge around 10000 cm$^{-1}$. However, 
below 3000 cm$^{-1}$ the reflectivity of the sample with $x$ = 
2.24 (top panel) increases steadily by lowering the temperature 
according to a conventional metallic behavior, while in the sample 
with $x$ = 1.56 (bottom panel) $R(\omega)$ has a more complicated 
behavior. Indeed, by lowering $T$ the reflectivity first increases 
down  to 250 K, then it decreases  to reach at 20 K its minimum 
value. 

\begin{figure}
{\hbox{\psfig{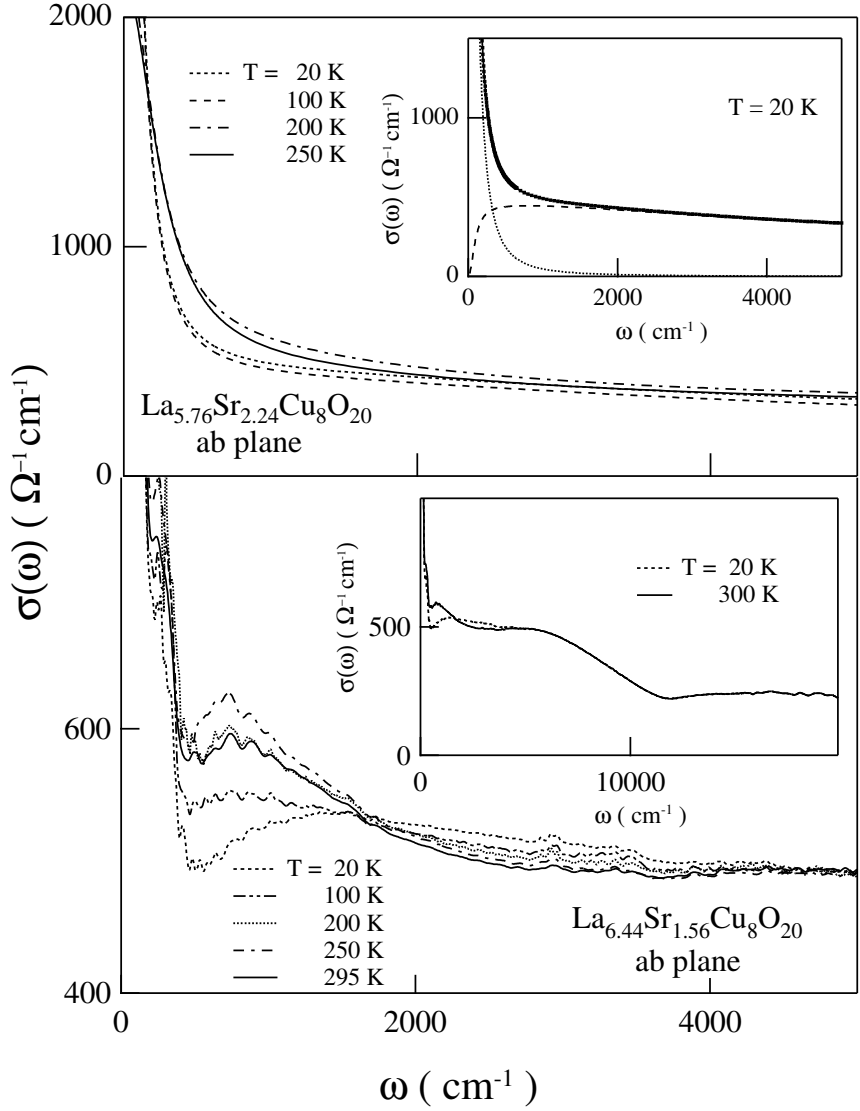}}}
\caption{Infrared optical conductivity in the $ab$ plane for the two 
crystals, as extracted from the $R(\omega)$ of Fig. 3 at different 
temperatures. In the upper inset, the experimental $\sigma(\omega)$ 
for $x$ = 2.24 is shown at 20 K (thick dots) together with a fit 
(solid line) given by a Drude term (dotted line) plus a broad 
background reproduced by a sum of Lorentzians (dashed line). In the 
lower inset, $\sigma(\omega)$ is shown for the $x$ = 1.56 crystal in 
the whole measuring range.}
\label{sigma-ab}
\end{figure}

The real part $\sigma(\omega)$ of the optical conductivity 
is shown in Fig.\ \ref{sigma-ab} for both samples, between 80 and 
5000 cm$^{-1}$ and at different temperatures. In the sample with $x$ 
= 2.24 (top panel) $\sigma(\omega)$ 
exhibits a pronounced metallic behavior. However, the behavior 
$\sigma(\omega) \propto \omega^{-1}$ that is observed in other 
cuprates and that inspired the "anomalous Drude" model of Eq.\ 
\ref{anomalous} is not observed here. On the other hand, a best 
fit to Eq.\ \ref{multicomp} gives very good results (see the inset, 
where the fit is superimposed to data) 
by using a normal Drude term plus a nearly flat background. This 
latter, in turn, is obtained by a superposition of Lorentzians peaked 
in the midinfrared (2500 cm$^{-1}$) and in the visible. The Drude 
plasma frequency ($\omega_D \sim$ 5500 cm$^{-1}$) is approximately 
independent of temperature, while $\Gamma_D$ decreases from 250 
cm$^{-1}$ at 295 K to 90 cm$^{-1}$ at 20 K. By using these values 
one predicts a $\sigma_{dc} = \omega_D^2/[60 \Gamma_D] \sim$ 6000 
(2000) $\Omega^{-1}$cm$^{-1}$ at 20 K (250 K) for the $x$ = 2.24 
sample, in agreement with the measured values\cite{Ito97} within a 
factor of 2.

In the sample with $x$ = 1.56 (bottom panel), 
where the anomalies in the dc properties are most evident, 
$\sigma(\omega)$ exhibits a complex structure. In the inset, 
$\sigma(\omega)$ is shown at two temperatures in the whole energy 
range. At least four 
absorption bands are directly observed in the spectrum, which then 
suggests a fit in terms of  the Drude-Lorentz approach. The band at 
the highest energy is a broad absorption located around 15000 cm$^{-
1}$, independent of $T$, which may be assigned
to the charge transfer transition between Cu and O. A second 
well-evident absorption is observed around 6000 cm$^{-1}$ and is 
nearly independent of temperature. This feature is then different 
from the midinfrared bands observed along the $c$ axis and 
assigned to bound charges. It is instead similar 
to the mid-infrared band (MIR) observed in most 
HCTS.\cite{Uchida,prb92} According to a pseudo-potential 
calculation\cite{Hermann} 
of the electronic structure of La$_4$BaCu$_5$O$_{13}$, a compound 
with lattice structure and transport properties similar to those 
of La$_{8-x}$Sr$_x$Cu$_8$O$_{20}$, the MIR band 
is due to a transition from an electronic band located at 
$\sim$ 0.5 eV below the Fermi energy $E_F$ to a band which 
crosses $E_F$. 

A third absorption in the bottom panel of Fig.\ \ref{sigma-ab} is 
strongly dependent on $T$ and extends from $\sim$ 500 to $\sim$ 4000 
cm$^{-1}$. Let us name its peak frequency $\omega_d$, for its 
similarity with 
other cuprates mentioned in the Introduction, where the 
corresponding band is assigned to the photoexcitation of polaronic 
charges.\cite{Falck,prl99,Lobo,Bucher}. At low temperature the $d$ peak 
is separated from the $T$-independent MIR band by a broad minimum 
centered at about 3000 cm$^{-1}$ (see also the inset). 
Another minimum at lower frequency, between this third absorption 
and the Drude term, deepens for decreasing temperature and 
recalls the optical 
pseudogaps reported for several underdoped cuprates.\cite{Timusk}
However, one sees from Fig.\ \ref{sigma-ab} that such deepening is due 
to a "blue shift" of the $d$ band. Indeed, in the present 
La$_{6.44}$Sr$_{1.56}$Cu$_8$O$_{20}$ crystal $\omega_d$ remains 
constant at $\sim$ 800 cm$^{-1}$ 
between 295 and 200 K, but at lower temperatures it {\it hardens} 
to reach about 1500 cm$^{-1}$ at 20 K. 
This behavior with temperature is quite different from that of 
superconducting cuprates, where usually $\omega_d$ {\it softens} for $T 
\to 0$. The intensity of the $d$ peak increases from 295 K to 250 K 
to decrease gradually from 250 to 20 K.

\begin{figure}
{\hbox{\psfig{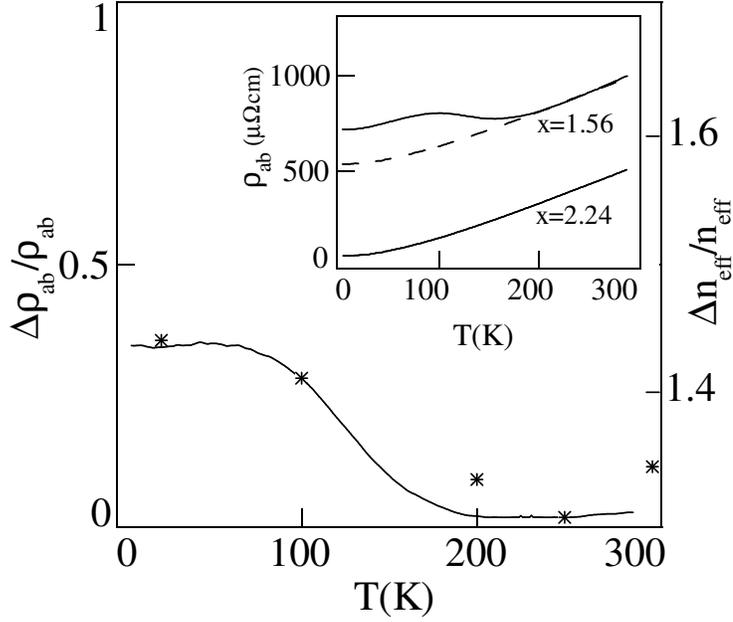}}}
\caption{Behavior with temperature of the infrared spectral weight 
of La$_{6.44}$Sr$_{1.56}$Cu$_8$O$_{20}$ in the $ab$-plane, compared 
with that of its dc 
resistivity $\rho_{ab}$. Raw $\rho_{ab}$ data are reported in 
the inset for both $x$ = 1,56 and $x$ = 2.24 as solid lines. The 
dashed line is the $\rho_{ab}$ of $x$ = 2.24, assumed as a normal 
metallic reference, once scaled by a constant factor in order to 
match the high-$T$ $\rho_{ab}$ of $x$ = 1.56. In the main Figure, 
the solid line gives $\Delta \rho_{ab} /\rho_{ab} = [\rho_{ab}(1.56) 
- \rho_{ab} (2.24)] / \rho_{ab}(2.24)$, namely the deviation of 
$\rho_{ab}$ in the $x$ = 1.56 sample from a normal metallic 
behavior. The stars represent the ratio $\Delta n_{eff}/n_{eff}$ = 
$[n_{eff}(80,5000)-n_{eff}(80,1700)]/n_{eff}(80,1700)$ as obtained 
from the $\sigma(\omega)$ of Fig. 4 through Eq. 4.}
\label{Delta-rho}
\end{figure}

One can easily check that the displacement of the $d$ band 
occurs via a transfer of spectral weight within the low-energy 
excitation spectrum of the carriers. Indeed, as one can see 
in the bottom panel of Fig.\ \ref{sigma-ab}, $\sigma(\omega)$ is 
independent of $T$  both at about 1700 cm$^{-1}$, and at 
5000 cm$^{-1}$. If one introduces the effective number of 
carriers

\begin{equation}
n_{eff}(\omega_1,\omega_2) = {2m^*V\over \pi e^2} 
\int_{\omega_1}^{\omega_2} \sigma(\omega) \, d \omega  \, .
\label{neff}
\end{equation}

\noindent
where $V$ is the cell volume and $m^*$ is assumed to be equal to the 
free electron mass, and puts $\omega_1 = 80$ cm$^{-1}$, 
$\omega_2$ = 5000 cm$^{-1}$, one obtains $n_{eff}(80,5000) = 
1.15 \pm 0.02$ 
at any $T$ from 295 K to 20K. Therefore, as $T$ changes, spectral weight 
is transferred across the fixed point at 1700 cm$^{-1}$ with no 
intervention from the electronic transitions at higher energies. 
This transfer is measured by the ratio $\Delta n_{eff}/n_{eff}$ = 
$[n_{eff}(80,5000)- n_{eff}(80,1700)]/n_{eff}(80,1700)$, that is 
reported at five 
temperatures in Fig.\ \ref{Delta-rho}. One should notice that the error
on this quantity is on the order of 2 \%. Therein, $\Delta 
n_{eff}/n_{eff}$ is also compared with the behavior of dc resistivity 
in the same sample. Indeed, in the inset of the Figure, the 
$ab$-plane raw resistivity $\rho_{ab}$ is reported by a solid line 
for both crystals here considered.\cite{Ito97}
Due to the reduced number of carriers, at $x$ = 1.56 $\rho_{ab}$ is 
larger than for the $x$ = 2.24 
sample, which behaves approximately like a conventional metal. 
However, if one scales $\rho_{ab}$ for this latter by a 
constant factor (dashed line), the two curves  show the 
same temperature dependence between 295 and 200 K. On the other 
hand, below 200 K, $\rho_{ab}$ for $x$ = 1.56 deviates significantly 
upwards from its behavior at $x$ = 2.24. In the main 
Figure, the solid line $\Delta \rho_{ab} /\rho_{ab} = 
[\rho_{ab}(1.56) - \rho_{ab} (2.24) )] / \rho_{ab}(2.24)$ measures 
such deviation. As one can see, this line qualitatively follows the 
transfer of infrared spectral weight towards higher frequencies as 
$T$ is lowered, indicated by $\Delta n_{eff}/n_{eff}$ (stars). 
Therefore, one finds that the variation in 
the dc resistivity is related to the "blue shift" of an infrared 
oscillator at finite frequency, well distinguished from the Drude 
term, the so-called $d$ band. This observation is consistent with 
the assignment of the $d$ band in this cuprate to self-trapped charges. 
Indeed, according to any polaronic model of $\sigma (\omega)$,\cite{Emin} 
$\omega_d \propto E_p$, 
the selftrapping energy of the carrier. Therefore, in this framework
the observed hardening of the $d$ band below 200 K 
implies a lower hopping rate 
of the carriers, consistent with the anomalous increase observed 
below 200 K in  the dc resistivity.

Finally, below 500 cm$^{-1}$ the $ab$-plane infrared conductivity 
of both samples in Fig.\ \ref{sigma-ab} shows a Drude-like free-
particle absorption. 
Phonon-like peaks are superimposed to it, more clearly for $x$ = 
1.56 due to a reduced screening effect of the carriers. The 
extrapolations of $\sigma(\omega)$ to $\omega$ = 0 at different 
temperatures agree within a factor of 2 with the corresponding 
values of $\sigma_{dc}$ extracted from the resistivities of Fig.\ 
\ref{sigma-ab}.

\section{Conclusion}

We have studied the optical properties of
La$_{8-x}$Sr$_x$Cu$_8$O$_{20}$, a non-superconducting cuprate 
characterized by a network of essentially one-dimensional paths. 
The two single crystals here examined have $x$ = 1.56 and $x$ = 
2.24, or approximately the lowest and highest doping, 
respectively, that have been studied in the literature for this 
compound.

We have first determined the optical response of the $c$ axis, where 
previous electron-diffraction experiments on La$_{8-
x}$Sr$_x$Cu$_8$O$_{20}$ showed clear superlattice spots for $x$ = 
1.6, diffuse superlattice scattering 
for $x$ = 2.24. In the optical spectra, for $x$ = 
1.56 we find a strong and $T$-dependent midinfrared band 
accompanied by a weak Drude term. At $x$ = 2.24 we observe a 
broad, nearly $T$-
independent, midinfrared background coexisting with a strong and 
narrow Drude term. The observation of such infrared features 
confirms that the cell doubling along the $c$ axis is due to 
charge ordering. Midinfrared bands similar to those reported here 
have been observed in other oxides (like nickelates and 
manganites) which exhibit either commensurate or incommensurate 
ordering. Therein, however, the Drude term is usually absent below 
the ordering temperature. In the present cuprate, on the contrary, 
the ordered charges coexist with carriers moving freely along the 
$c$ axis, because the two species are probably placed on different 
paths. In any case, once 
again the observation of a $T$-dependent midinfrared 
absorption in a cuprate is intimately related to selftrapped, or 
polaronic, charges, that at high doping may form ordered 
structures.  

We have then studied the $ab$ plane of 
La$_{8-x}$Sr$_x$Cu$_8$O$_{20}$, where no 
indications of superlattice features are extracted from the 
diffraction experiments. The infrared spectra of 
the $x$ = 1.56 crystal exhibit again both a Drude-like term and a 
well resolved midinfrared absorption. However the latter band is 
much softer than for the $c$ axis of the same sample ($\sim$ 0.1 
eV with respect to 0.3 eV). A fit to the $x$ = 2.24 optical 
conductivity confirms the above two-term scenario also for the 
highest doping. However, unlike along the $c$ axis, in the $ab$ 
plane one could have a single type of lightly bound carriers, 
possibly large polarons. The Drude part of the spectrum would 
reflect their behavior as quasi-free particles, the soft band at 
0.1 eV would correspond to their photoionization, {\it i. e.}, their 
destruction by absorption of a photon. This interpretation 
is consistent with the observed correlation between the $T$-
dependence of the band at 0.1 eV and that of the dc resistivity in 
the same $x$ = 1.56 sample.

By recalling what reported in the Introduction, the $ab$ 
plane of La$_{8-x}$Sr$_x$Cu$_8$O$_{20}$, behaves in the infrared 
like the $ab$ plane of the superconducting cuprates, even if it has 
a quite 
different geometrical structure. However, a crucial difference has 
been pointed out by the present experiment. While in the non-
superconducting 
8-8-20 the infrared bands related to the bound charges move to 
higher frequencies for $T \to 0$, indicating stronger 
localization, in some high-$T_c$ superconductors the corresponding 
absorption is  found to soften for decreasing temperature in the 
normal phase, indicating a carrier delocalization or - in a phase 
separation scenario - stronger charge-density fluctuations. This 
result should be taken into consideration by those models where 
the existence of bound charges is related to the still unexplained 
phenomenon of high-$T_c$ superconductivity. 

\acknowledgements

We are indebted to M. Capizzi and P. Quemerais for useful 
discussions and suggestions.

\newpage


\end{document}